\documentclass[preprint2,10pt]{aastex}

\usepackage{amsmath}
\usepackage{color}

\newcommand{\prep}[1]{#1}
\newcommand{\manu}[1]{}

\prep{\setlength\textwidth{18cm}}
\prep{\setlength\textheight{24.5cm}}
\prep{\setlength\columnsep{0.5cm}}
\prep{\setlength\topmargin{-2.0cm}}
\prep{\hoffset =-1.5cm }
\prep{\renewcommand{\baselinestretch}{0.92}}

\def\abs#1{\left\vert#1\right\vert}

\def\dop#1#2{{#1}\cdot{#2}}
\def\trans#1{{}^t{#1}}

\newcommand{\te}[1]{\times10^{#1}}

\newcommand{\secyear}{\arcsec\!\cdot{\rm{yr}}^{-1}}

\newcommand{\bn}{\mathbf{n}}

\newcommand{\bw}{\mathbf{w}}
\newcommand{\bi}{\mathbf{i}}
\newcommand{\bj}{\mathbf{j}}
\newcommand{\bk}{\mathbf{k}}
\newcommand{\vmax}{{\Theta_{\rm max}}}
\newcommand{\omegal}{{\omega_{\rm lib}}}

\def\crm{\cr\noalign{\medskip}}

\def\m@th{\mathsurround=0pt}
\def\EQM#1{\vcenter{\normalbaselines\m@th
    \ialign{${\displaystyle ##}$\hfil&&\ ${\displaystyle ##}$\hfil\crcr
    \mathstrut\crcr\noalign{\kern-\baselineskip}
    \noalign{\smallskip}
    #1\crcr\mathstrut\crcr\noalign{\kern-\baselineskip}}}}

\newcommand{\be}{\begin{equation}}
\newcommand{\ee}{\end{equation}}

\def\grad#1#2{{\boldsymbol \nabla}_{#2}{#1}}

\newcommand{\bpm}{\begin{pmatrix}}
\newcommand{\epm}{\end{pmatrix}}

\newcommand{\corr}[1]{{#1}}

\newcommand\figa{
\begin{figure}[t]
\begin{center}
\includegraphics[width=0.5\linewidth]{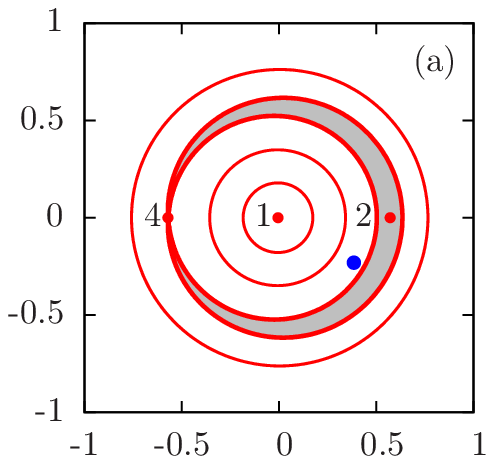}\includegraphics[width=0.5\linewidth]{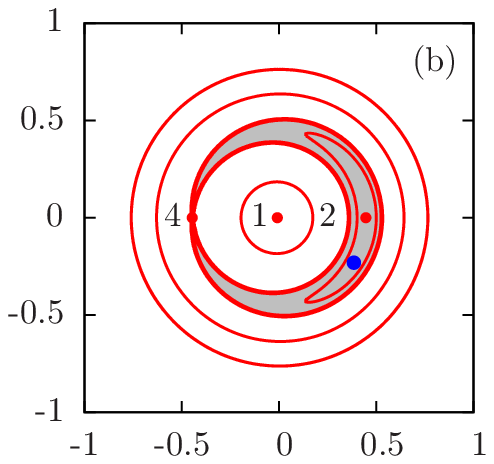}
\figcaption{Projection of the spin axis $\bw$ (\ref{eq.wcoord}) in the orbital frame
(in abscissae $\sin\epsilon\cos\psi$, in ordinate $\sin\epsilon\sin\psi$).  (a) Case I with
$\alpha=0.845\secyear$ \citep{Helled_etal_Icarus_2009}. (b) 
Case II with $\alpha=0.775\secyear$ \citep{Ward_Hamilton_AJ_2004}.
Cassini state 3 corresponds to a retrograde rotation of Saturn and is
not represented in these figures. The current position of Saturn's spin
axis is represented by a large filled circle. The small filled circles
are Cassini states and the curves are energy contours. The bold curve is
the separatrix that delineates the libration area in grey.\label{fig.cass}}
\end{center}
\end{figure}
}

\newcommand\figb{
\begin{figure}[t]
\begin{center}
\prep{\includegraphics[width=\linewidth]{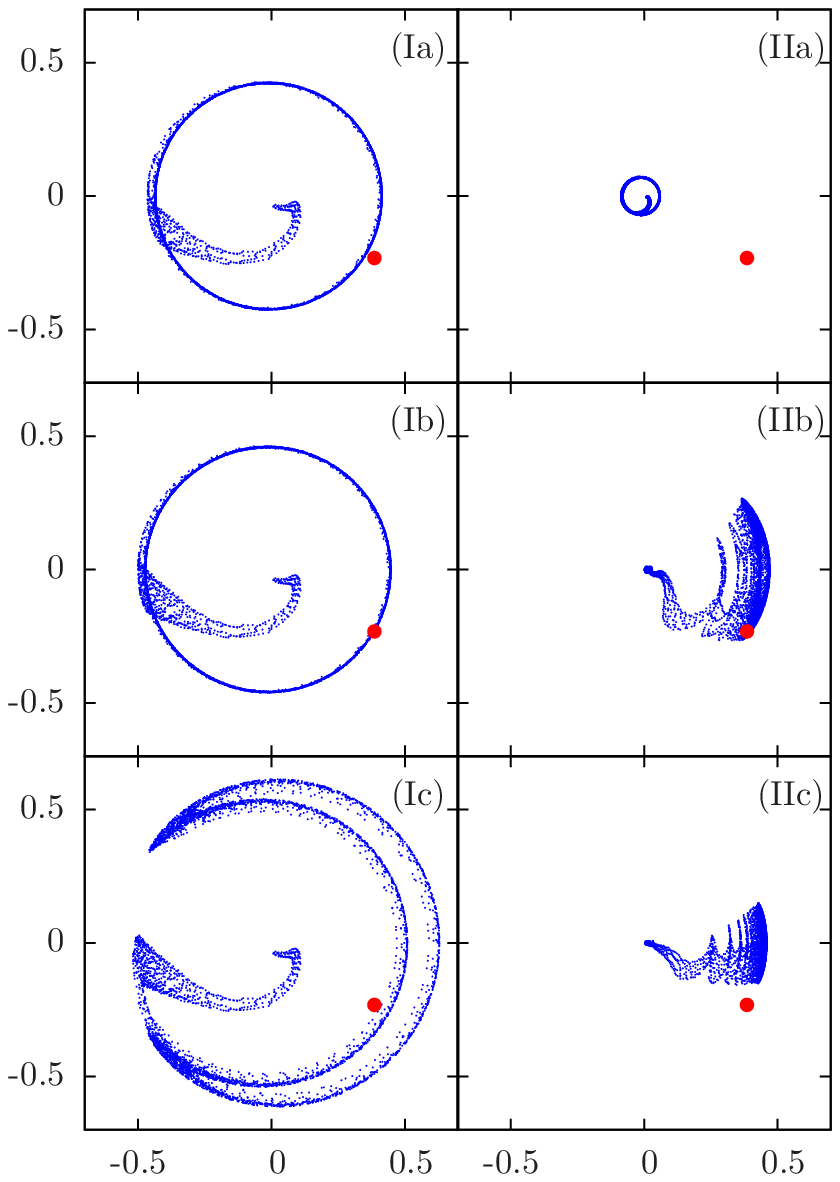}}
\manu{\includegraphics[width=0.5\linewidth]{BLK_fig2.ps}}
\figcaption{Results for the MA type migration.
Projection of Saturn's spin axis on the invariant plane in a frame
rotating at Neptune regression frequency (in abscisse $\sin\theta\cos(\phi-\Omega)$, 
in ordinate $\sin\theta\sin(\phi-\Omega)$) (see Eq.~\ref{eq.wcoordi}). The filled circle represents its
current position. Subfig. Ia, Ib and Ic, Case I
with \corr{$\tau=180$ Myr} and $\psi=108.674, 108.675\,469,
108.675\,475\deg$. Subfig. IIa, IIb and IIc, Case II with \corr{$\tau = 20,
200, 300$ Myr} and $\psi=0\deg$.\label{fig.evol}}
\end{center}
\end{figure}
}

\newcommand\figd{
\begin{figure}[t]
\begin{center}
\includegraphics[width=0.75\linewidth]{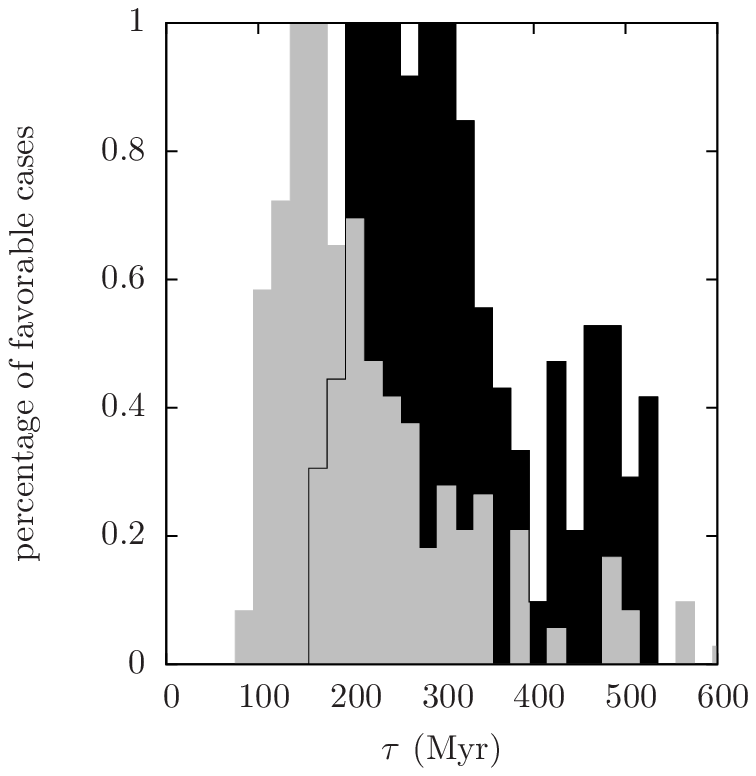}
\figcaption{Probability that Saturn librates in zone 2 with an amplitude larger 
than $31\deg$ as a function of the migration time scale $\tau$ in Case II. MA migration
type in grey and DE migration type in black.
\label{fig.histo}}
\end{center}
\end{figure}
}

\newcommand\fige{
\begin{figure}[t]
\begin{center}
\includegraphics[width=\linewidth]{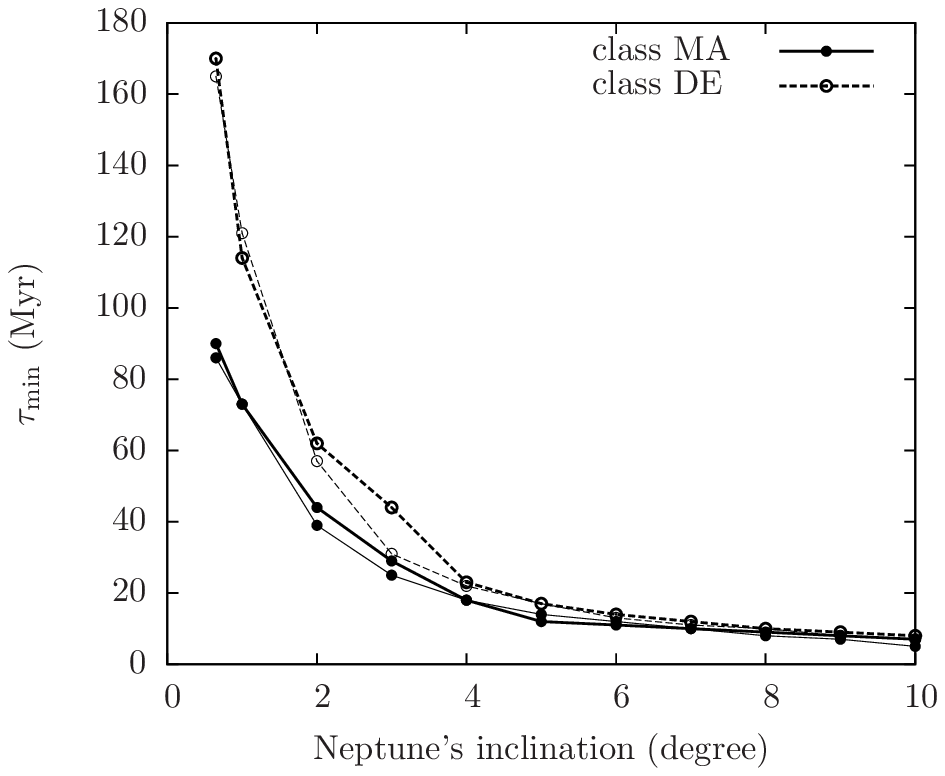}
\figcaption{Minimal migration time scale as a function of Neptune's
initial inclination for both migration types: MA (solid line) and DE (dashed line).
Bold curves are results of numerical integrations. Thin curves were
obtained by the algorithm described at the end of Section~\ref{sec.anal}.
\label{fig.taumin}}
\end{center}
\end{figure}
}

\newcommand\tabb{
\begin{table}[h]
\begin{center}
\caption{\label{TTb}Secular Frequencies Associated with the Precession of
the Ascending Nodes.}
\begin{tabular*}{\linewidth}{@{\extracolsep{\fill}}lrr}\hline
Frequency  &  Laskar et al. (2004) &  Secular integration \\ \hline
$s_5$      &  -0.000        &  -0.000        \\
$s_6$      & -26.348        & -26.569        \\
$s_7$      &  -2.993        &  -2.996        \\
$s_8$      &  -0.692        &  -0.689        \\ \hline
\end{tabular*}
\end{center}
\end{table}
}

\shorttitle{Tilting Saturn}
\shortauthors{Bou\'e \& Laskar}

\begin{document}
\slugcomment{{\sc \small The Astronomical Journal}}

\title{\prep{\uppercase}{Speed limit on Neptune migration imposed by 
Saturn tilting}}

\author{\prep{\sc} Gwena\"el Bou\'e, Jacques Laskar and Petr Kuchynka}
\affil{Astronomie et Syst\`emes Dynamiques, IMCCE-CNRS UMR8028,
Observatoire de Paris, UPMC, 77 Av. Denfert-Rochereau, 75014 Paris, France}
\email{boue@imcce.fr}

\date{\today}

\begin{abstract}
In this Letter, we give new constraints on planet migration. They were
obtained under the assumption that Saturn's current obliquity is due to
a capture in resonance with Neptune's ascending node. 
If planet migration
is too fast, then Saturn crosses the resonance without being captured
and it keeps a small obliquity. This scenario thus gives a lower limit
on the migration time scale $\tau$. We found that this boundary depends
strongly on Neptune's initial inclination. For two different migration
types, we found that $\tau$ should be at least greater than 7 Myr. This
limit increases rapidly as Neptune's initial inclination decreases from
10 to 1 degree. We also give an algorithm to know if Saturn can be tilted
for any migration law.
\end{abstract}


\section{Introduction}

It is now well accepted that the Solar System was more compact after the
protoplanetary gas disk dissipated. Then planets migrated due to
interactions with the primordial planetesimal belt. The Nice model
\citep{Gomes_etal_Nature_2005, Tsiganis_etal_Nature_2005,
Morbidelli_etal_Nature_2005} gives a unified scenario of this planetary
migration, but it is still not fully constrained. For example, the Nice
model allowed two possible classes of late evolution 
\citep{Nesvorny_etal_AJ_2007}. In the first one, called ``class
MA'', Neptune is scattered to 22-25 AU
and reaches its final orbit by slowly migrating over more than 5 AU. In
the second class, labeled ``DE'',
Neptune is placed to its current orbital distance with large
eccentricity $\approx0.3$ and then slowly circularizes. Besides, the
Nice model does not constrain inclinations, and the time scale of this
late evolution is uncertain. Nevertheless, constraints on the migration time scale
were obtained from the distribution of the Kuiper belt on the one hand
\citep{Murray-Clay_Chiang_ApJ_2005}, and from the distribution of the main
asteroid belt on the other \citep{Minton_Malhotra_Nature_2009}.
They both assumed an MA migration type without long-term evolution of
eccentricities and inclinations. The former obtained a migration time
scale $\tau$ between 1 and 10 Myr, and the latter found $\tau\lesssim0.5$ 
Myr.

In this Letter, we aim at giving new constraints based on Saturn tilting
\citep{Ward_Hamilton_AJ_2004, Hamilton_Ward_AJ_2004}. According to Ward
and Hamilton, Saturn's large obliquity, $\epsilon=26.73919\deg$
\citep{Helled_etal_Icarus_2009}, is due to a resonance capture between
its spin axis and Neptune's orbit.
Given the large uncertainties on Saturn's precession rate
$-0.75\pm0.21\secyear$ \citep{Ward_Hamilton_AJ_2004}, the more accurate
regression of Neptune's orbit plane $s_8=-0.692\secyear$
\citep{Laskar_etal_AA_2004} is indeed included in the errorbars.
\citet{Ward_Hamilton_AJ_2004} assume that today the two frequencies
are equal.
In their scenario, the norm of the frequency of Neptune's ascending node
was initially larger, and then it captured Saturn's spin 
axis as it decreased due to Neptune's migration and/or the 
dissipation of the planetesimal disk \citep{Ward_Canup_ApJ_2006}. In their
numerical model, they
took a quasiperiodic model of the Solar System and forced an exponential
evolution of the frequency $s_8$. Here, we show that Saturn can
tilt in both migration classes, and that it gives a lower limit on
the migration time scale. This limit depends on Neptune's initial
inclination.

A recent paper by \citet{Helled_etal_Icarus_2009} seems to contradict
Ward and Hamilton's scenario. It gives a new estimate of Saturn's
precession rate $-0.7542\pm0.0002\secyear$ that is incompatible with 
a resonance with $s_8$. We show that with this value, Saturn can still
evolve to its current state but that it is very unlikely. We discuss this
result in our conclusion.

\section{Spin axis evolution}
\label{sec.evol}
Here we recall the equations of motion of a planet axis and give the
current dynamical state of Saturn's spin axis.
The evolution of the spin axis $\bw$ of a planet in a fixed reference
frame $(\bi,\bj,\bk)$ where $\bk$ is the direction of the total orbital 
angular momentum,  is given by
\be
\frac{d\bw}{dt} = -\alpha(\dop{\bn}{\bw})\bn\times\bw
\label{eq.evols}
\ee
where $\bn=\trans{(n_x,n_y,n_z)}$ is the normal to the orbit, and $\alpha$
is the precession constant. Without planetary perturbations, $\bn$
is fixed and the spin axis $\bw$ precesses uniformly around $\bn$ with
constant obliquity $\cos\epsilon=\dop{\bn}{\bw}$. However, in a multi-planetary
system, $\bn$ evolves due to secular interactions. 
The long-term evolution of $\bn$ can be approximated
by a quasiperiodic expression \be
n_x+in_y = \sum_k I_k e^{i(\nu_k t+\varphi_k)}\ ,
\label{eq.anaf}
\ee
where the  $\nu_k$ (sorted with increasing amplitudes $I_k$) are combinations of the fundamental
frequencies $g_j$, $s_j$ \citep{Laskar_1990}.
For Saturn, $\nu_2=s_8=-0.692\secyear$ and
$I_2=0.064\deg$.
As the other terms have
only very weak effects on the behavior of Saturn's spin axis
\citep{Hamilton_Ward_AJ_2004},  
one can retain this single term in the orbital precession, which makes the problem integrable
 \citep{Colombo_AJ_1966,Henrard_CeMe_1987}.
The associated autonomous Hamiltonian, written in a moving reference frame 
related to the orbital plane, reads
\be
H=-\frac{\alpha}{2}(\dop{\bn}{\bw})^2-\nu(\dop{\bk}{\bw})
\ee
with now $\bk=\trans{(I,0,\sqrt{1-I^2})}$ and $\bn=\trans{(0,0,1)}$. The
equation of motion, obtained from ${d\bw}/{dt}=\grad{H}{\bw}\times\bw$,
is
\be
\frac{d\bw}{dt} = -\alpha(\dop{\bn}{\bw})\bn\times\bw-\nu\bk\times\bw\ .
\ee
This system possesses 4 relative equilibriums named Cassini states for which
 the three axes $\bw$, $\bn$ and $\bk$ are collinear, and a
separatrix delineating 3 zones in the phase space (see
Fig.~\ref{fig.cass}). Hereafter, we label the 3 zones after the Cassini
state they contain. Saturn's spin axis coordinates in the orbital frame
are 
\be
\bw=\bpm \sin\epsilon\cos\psi \\ \sin\epsilon\sin\psi \\ \cos\epsilon \epm
\label{eq.wcoord}
\ee
with $\epsilon=26.73919\deg$ \citep{Helled_etal_Icarus_2009} and $\psi=-31\deg$
\citep{Hamilton_Ward_AJ_2004}. 
As $\psi\neq0$, the system is not in a Cassini state. Given the
coordinates of $\bw$, if $-\alpha\cos\epsilon\in[-0.730,-0.666]$ $\secyear$ 
\citep{Ward_Hamilton_AJ_2004} then Saturn's spin axis is in resonance
around Cassini state 2 with a libration amplitude larger than $31\deg$
(see Fig.\ref{fig.cass}b), else it is in circulation around either 
Cassini state 1 (see Fig.\ref{fig.cass}a) or Cassini state 3.
Literature gives three different values of Saturn's precession rate. Two
of them, $-0.74\pm0.7\secyear$ \citep{French_etal_Icarus_1993} and
$-0.75\pm0.21\secyear$ \citep{Ward_Hamilton_AJ_2004} are compatible with
a libration in zone 2, whereas the third one, $-0.7542\pm0.0002\secyear$
\citep{Helled_etal_Icarus_2009}, constrains Saturn's axis to circulate
in zone 1. In the following, we study these two cases. In Case I, we use
the precession constant given by \citet{Helled_etal_Icarus_2009}, and in
Case II we set $\alpha$ such that $-\alpha\cos\epsilon=s_8=-0.692\secyear$.
In our numerical integrations detailed below, we take into account the
dependences of $\alpha$ in Saturn's semi-major axis and eccentricity. 

\prep{\figa}

\section{Orbital evolution}
\label{num}
We integrate the secular equations of motion derived from the
Hamiltonian of \citet{Laskar_Robutel_CeMe_1995} written up to
degree 4 in inclinations and eccentricities. 
In order to fit  to the present value of $s_8$ \citep{Laskar_etal_AA_2004}, 
a small constant offset
 $\delta s_8 = -0.00342\secyear$ is added in the model. This offset was obtained  
 by frequency
analysis \citep{Laskar_1990} of our analytical model (Table \ref{TTb}). 
For the class ``MA'', we consider only the last 3 AU migration of
Neptune. When Neptune was closer to the Sun, the frequency $s_8$ was too
large to have any effect on Saturn's axis. Migration is simulated by an
additional force leading to the following exponential law,
\be
a(t) = a_0+\Delta a(1-e^{-t/\tau}),
\ee
with $\Delta a=+0.1, -0.3, -1.3, -3$ AU respectively for Jupiter, Saturn,
Uranus, and Neptune. It is scaled from \citet{Minton_Malhotra_Nature_2009}
and it is in agreement with the full integration of 
\citet{Tsiganis_etal_Nature_2005}. In the same way, for the class ``DE''
we apply an external force that gives a long-term exponential evolution
of Neptune eccentricity starting at $e_0=0.3$ and finishing at its current
value. For both classes, we did integrations with constant Neptune
inclination, and others with an exponential damping with the same $\tau$. 
For each value of $\tau$,
an integration in the past is done to obtain initial conditions
for the orbital coordinates. Saturn's initial obliquity is then set to
$\epsilon_0=1.5\deg$.

\prep{\tabb}

\prep{\figb}

We now look at the effect of the dissipation of the remaining primordial
planetesimal belt.
Following a suggestion of \citet{Morbidelli_Comm_2009},
the mass $m_K$ of the planetesimal belt in the class ``MA''
is estimated by energy conservation as follows.
Initially, planetesimals are distributed following
\citet[][fig. 1]{Morbidelli_etal_MNRAS_2004} and during Neptune migration,
planetesimals move from their initial position to Uranus' orbit. This leads
to $m_K = 1.7\pm0.1M_\oplus$.  During planet migration with a planetesimal
disk, we force an exponential decrease of the planetesimal belt mass with
the same time scale as the semi-major axis one. To first order, 
the averaged effect of a planetesimal of mass $m_i$ and semi-major axis
$a_i$ on Neptune's nodal precession rate is 
\be
\delta s_8(m_i,a_i) = -\frac{n_N}{4}\left(\frac{m_i}{m_0}\right)
\left(\frac{a_N}{a_i}\right)^2 b_{3/2}^{(1)}(a_N/a_i)
\ee
with $n_N$ and $a_N$ being respectively Neptune mean motion and semi-major
axis, and $m_0$ the mass of the Sun. We model the planetesimal belt by a
single annulus with semi-major axis $a_K$ such that 
\be
\delta s_8(m_k,a_k)=\sum_i \delta s_8(m_i,a_i)\ .
\ee
Using the mass distribution of \citet{Morbidelli_etal_MNRAS_2004}, we found
$a_k=60\pm5$AU. We run numerical integrations with and without a
2 Earth mass primordial planetesimal belt and found that the constraints
on the migration time scales were unchanged. In the following, we give
only the results of our integrations without a planetesimal disk.

\prep{\figd}

\section{Results}
\label{sec.results}

In Case I, Saturn's spin axis circulates around Cassini state 1
with a large obliquity (Fig.~\ref{fig.cass}a). To show whether it is
compatible with the \citet{Ward_Hamilton_AJ_2004} scenario or not, we did several
numerical integrations without long-term evolution of Neptune inclination 
and enumerated those ending in zone 1
with an obliquity larger than or equal to $26.73919\deg$. We first considered
the class MA of the Nice model and varied the migration time scale
$\tau$ from 100 Myr to 600 Myr every 10 Myr. Then, for each value of
$\tau$, we searched the range of the initial precession angle $\psi$ for
which the final state corresponds to our criterion. We found only 3
values for $\tau$ satisfying the criterion: $\tau\in\{170,180,190\}$
Myr. In each case, the range of possible values for $\psi$ is extremely
small $\Delta_\psi\lesssim10^{-5}\deg$ (Fig.~\ref{fig.evol}, Iabc). Thus
assuming an equiprobable initial phase, the probability to
find Saturn in its current state through this mechanism is less than
$3\times10^{-8}$ for any of the three selected $\tau$. With a DE migration
type, the widths of the initial longitude intervals $\Delta_\psi$ are
identical. The only changes are in the values of the migration time scale
$\tau$ leading to the large obliquity circulation state: $\tau\in\{150, 
260, 290, 310, 320\}$ Myr.
We discuss the implications of these results in the conclusion.

In Case II, Saturn spin axis is presently in
resonance with Neptune's ascending node. In that case, planet migration
must be slow enough for the capture to occur, but if it is too slow, then the
evolution becomes adiabatic and the
libration amplitude is too small (less than $31\deg$) (Fig~\ref{fig.evol}, IIc). 
This latter constraint disappears if the precapture obliquity is larger
than $4.5\deg$ \citep{Ward_Hamilton_AJ_2004}.
We performed 2100 integrations for each of the two migration types MA and
DE, $\tau$ going from 10 to 600 Myr every $10$ Myr and $\psi$ between 
$0$ and $350\deg$ every
$10\deg$. The results are summarized in Fig.~\ref{fig.histo}, MA type in
grey and DE type in black. Probabilities are now significant and reach 1
for a few time scales. We see clear lower limits, $\tau\geq90$ Myr
(resp. $\tau\geq170$ Myr) for the MA (resp. DE) migration type.
The difference in the results between the two Nice model classes comes
mainly from the different dependence of the semi-major axis and the
eccentricity on Neptune's regression frequency.
In all these integrations, Neptune's inclination does not undergo
long-term evolution. However, the amplitude $I_2$ of Saturn orbital
quasiperiodic motion (\ref{eq.anaf}) is proportional to Neptune
inclination. In Section~\ref{sec.anal}, we show that the higher the
inclination amplitude is, the faster a planet can be tilted. We thus studied 
the minimum time scale, for which Saturn's axis ends in zone 2 with a
libration amplitude larger than $31\deg$, as a function of Neptune's
initial inclination (Fig.~\ref{fig.taumin}, bold curves).
In both migration classes, $\tau_{\rm min}$ decreases rapidly to
$\approx20$ Myr when Neptune's initial inclination increases to $4\deg$ 
and then it decreases slowly down to $\approx7$ Myr when Neptune's
inclination goes to $10\deg$.

\prep{\fige}

\section{Fastest tilting}
\label{sec.anal}
In this section, we compute analytically the minimal time required to
tilt a planet as a function of its inclination $I(t)$. We give also an 
algorithm to check whether Saturn can be tilted or not for a given migration.
 
We call $\theta$ the inclination of a planet equator relative to the
invariant plane. As Saturn current inclination is small relative to its
obliquity $\epsilon$, the two angles $\theta$ and $\epsilon$ are similar.
Let $\Phi$ and $\Omega$ be the longitude of the ascending node of the
equator and of the orbit in the invariant plane. We have
\be
\bw=\bpm\sin\theta\sin\Phi \\ -\sin\theta\cos\Phi \\
\cos\theta\epm_{(\bi,\bj,\bk)}
\quad
\bn=\bpm\sin I\sin\Omega \\ -\sin I\cos\Omega \\ \cos
I\epm_{(\bi,\bj,\bk)}
\label{eq.wcoordi}
\ee
and 
\be
\EQM{
\dop{\bn}{\bw} &=& \sin\theta\sin I\cos(\Phi-\Omega)+\cos\theta\cos I,
\crm
\dop{(\bn\times\bw)}{\bk} &=& -\sin\theta\sin I\sin(\Phi-\Omega).
}
\ee
Thus, from (\ref{eq.evols}),
\be
\frac{d\cos\theta}{dt} 
= \alpha\sin\theta\cos\theta\sin I\cos I 
[1+\gamma\cos(\Phi-\Omega)]\sin(\Phi-\Omega)
\label{eq.dcosdt}
\ee
where $\gamma=\tan\theta\tan I$  can vary from 0 to infinity
depending on the value of the obliquity. We now choose $\Omega$
that maximizes this time derivative as a function of $I$ and $\theta$.
Doing so, we ensure that it is not possible to have a
faster evolution of the equator inclination $\theta$. This leads to
\be
\cos(\Phi-\Omega)=\frac{-1+\sqrt{1+8\gamma^2}}{4\gamma}.
\ee
Substituting this expression in (\ref{eq.dcosdt}) gives the maximal speed
$\vmax$ such that $d\theta/dt\leq\vmax$
\be
\vmax=\frac{\alpha\sqrt{2}}{16}\abs{\cos\theta\sin 2I}
\frac{\left(\sqrt{1+8\gamma^2}+3\right)^{3/2}}{\left(\sqrt{1+8\gamma^2}+1\right)^{1/2}}.
\label{eq.fastest}
\ee

After some calculus, it can be shown that $\vmax$ is an increasing
function of $\tan I$. Thus, if the only constraint on
the orbit inclination amplitude is an upper limit $I_{\rm max}<\pi/2$, the
fastest evolution is obtained for $I=I_{\rm max}$. In two asymptotic cases, the
expressions of $\vmax$ are simpler. For $I\ll\abs{\pi/2-\theta}$ or $\theta\ll\abs{\pi/2-I}$,
we have $\abs{\gamma}\ll1$ and thus
\be
\vmax\approx\frac{\alpha}{2}\abs{\cos\theta{\sin2I}}.
\label{eq.approx}
\ee
In the other case, if $\abs{\theta-\pi/2}\ll I$ or $\abs{I-\pi/2}\ll \theta$, the
parameter $\gamma$ is arbitrarily large and (\ref{eq.fastest}) becomes
\be
\vmax\approx\frac{\alpha}{2}\sin\theta\sin^2I.
\ee
Using the approximation for small angles $\abs{\gamma}\ll1$ (\ref{eq.approx}),
the minimum time \corr{$t_{\rm min}$} required to bring
$\theta$ from 0 to $\theta_{\rm end}$ at constant inclination amplitude $I$ is
\be
\corr{t_{\rm min}} = -\frac{2}{\alpha\sin 2I}
\ln\abs{\tan\left(\frac{\pi}{4}-\frac{\theta_{\rm end}}{2}\right)}.
\label{eq.tmin}
\ee
In Saturn's case, the amplitude of the mode responsible for the
tilt is $I_2$. Whenever Neptune's inclination is less than $10\deg$, $I_2$
remains below $0.9\deg$, $\gamma\leq8\te{-3}$ and Expressions
(\ref{eq.approx}, \ref{eq.tmin}) are valid. This
minimum time $t_{\rm min}$ decreases with the inclination
amplitude $I$. For example, with $I=I_2=0.064\deg$, Equation (\ref{eq.tmin})
gives $t_{\rm min}=105$ Myr in Case I and $t_{\rm min}=115$ Myr in Case 2.

From this study, it is possible to check whether Saturn's axis can be tilted or not 
for a given migration. Let $\theta_2$ be the value of $\theta$ at the Cassini
state 2. $\theta_2$ and its time derivative $\Theta_2=\dot{\theta}_2$
are functions of orbital parameters through $I(t)$, $\nu(t)=s_8(t)$ and
$\alpha(t)$. During a tilt, $\theta$ oscillates around the increasing 
$\theta_2$ and $\Theta=\dot{\theta}$ reads
\be
\Theta = \Theta_2 + A\omegal\sin(\omegal t+\varphi)
\label{eq.Theta}
\ee
where $A$ and $\varphi$ are respectively the libration amplitude and a
phase, and $\omegal$ is the libration amplitude given by
\citet{Hamilton_Ward_AJ_2004}
\be
\omegal = \sqrt{-\alpha\nu\sin\theta\sin I}.
\label{eq.omegal}
\ee
For a given migration, Saturn's axis can tilt if and only if there exist
$A$ and $\phi$ such that $\Theta\leq\vmax$ (\ref{eq.approx} and \ref{eq.Theta}) 
during all the evolution. Replacing $\theta$ by $\theta_2$ in
(\ref{eq.approx} and \ref{eq.omegal}), one obtains a criterion that
depends only on orbital parameters.
We applied this criterion on the systems studied in
Section~\ref{sec.results}. For each value of Neptune's initial inclination, we 
integrated once the system with a given $\tau$. Then, we rescaled the
derivatives $\Theta$ for different value of $\tau$ until the criterion
is verified. The resulting values of $\tau_{\rm min}$ are displayed in
Fig. (\ref{fig.taumin}, thin curves).

\section{Conclusions}
First of all, we see that the \citet{Helled_etal_Icarus_2009} precession constant
is incompatible with the \citet{Ward_Hamilton_AJ_2004} scenario. This is a
robust result.
\citet{Helled_etal_Icarus_2009} obtained Saturn's precession constant from an
empirical model of its internal structure.
They used Saturn mass, radius and gravitational coefficients $J_2$,
$J_4$, and $J_6$ to fit a density profile represented by a sixth degree
polynomial. From this density profile they derived the normalized axial moment of
inertia $\gamma$ directly related to the precession constant. Our results 
suggest, rather, considering $\gamma$ as an additional independent parameter to
better constrain Saturn's interior. If Saturn is actually in libration in zone 2
then $0.2257<\gamma<0.2438$ \citep{Ward_Hamilton_AJ_2004}.

Assuming the \citet{Hamilton_Ward_AJ_2004} precession constant, Saturn's spin
axis is likely to evolve toward a libration in zone 2 whatever the
migration class is as long as the time scale $\tau$ is sufficiently
large. We found a strong dependence between the minimum time scale $\tau_{\rm
min}$ and Neptune's inclination. Thus, an external
constraint on the speed limit of Neptune migration may also constrain
its inclination.
For instance, the upper boundary obtained by
\citet{Murray-Clay_Chiang_ApJ_2005} is $\tau\leq10$ Myr. In that case,
our results show that under the hypothesis of Section~\ref{num}, the
initial inclination of Neptune's orbit must have been larger than 7 deg.
On the other side, in all our studied cases, the minimum time scale must
be at least greater than 7 Myr, whereas \citet{Minton_Malhotra_Nature_2009}
found $\tau\lesssim0.5$ Myr. This contradiction may be raised if one
considers different evolution laws for the semi-major axes, eccentricities,
and/or inclinations. In that scope, we have given in
Section~\ref{sec.anal} an algorithm to know if Saturn can be tilted for
any migration law.

\acknowledgments


\prep{\small}
\prep{\renewcommand{\baselinestretch}{0.8}}

\prep{\end{document}}


\clearpage

\figa

\figb

\figd

\fige


\clearpage

\tabb

\end{document}